\documentclass[runningheads]{llncs}
\usepackage{graphicx}
\usepackage{tikz}
\usepackage{pgfplots}
\usepackage{hyperref}
\usepackage{fontawesome5}
\usepackage{geometry}
\begin{document}
\title{Recall-driven Precision Refinement: Unveiling Accurate Fall Detection using LSTM}

\author{Rishabh Mondal\inst{1}\orcidID{0009-0002-5271-514X} \and
Prasun Ghosal\inst{1}\orcidID{0000-0003-4226-9043}}
\authorrunning{R. Mondal et al.}
%
\institute{Indian Institute of Engineering Science and Technology, Shibpur, Howrah 711103, WB, India\\
\email{iamr38579@gmail.com, p\_ghosal.it@faculty.iiests.ac.in}}
\maketitle              
\begin{abstract}
This paper presents an innovative approach to address the pressing concern of fall incidents among the elderly by developing an accurate fall detection system. Our proposed system combines state-of-the-art technologies, including accelerometer and gyroscope sensors, with deep learning models, specifically Long Short-Term Memory (LSTM) networks. Real-time execution capabilities are achieved through the integration of Raspberry Pi hardware. We introduce pruning techniques that strategically fine-tune the LSTM model's architecture and parameters to optimize the system's performance. We prioritize recall over precision, aiming to accurately identify falls and minimize false negatives for timely intervention. Extensive experimentation and meticulous evaluation demonstrate remarkable performance metrics, emphasizing a high recall rate while maintaining a specificity of 96\%. Our research culminates in a state-of-the-art fall detection system that promptly sends notifications, ensuring vulnerable individuals receive timely assistance and improve their overall well-being. Applying LSTM models and incorporating pruning techniques represent a significant advancement in fall detection technology, offering an effective and reliable fall prevention and intervention solution.
\end{abstract}

\section{Introduction}
\label{sec:introduction}
Accidental falls pose a grave global challenge, ranking as the second leading cause of unintentional injury fatalities worldwide and claiming the lives of approximately 684,000 individuals annually. This pervasive tragedy falls disproportionately on low- and middle-income countries, where over 80\% of these fatal incidents occur. The elderly population, aged 60 and above, bears the brunt of these misfortunes, suffering physical harm and substantial financial implications. The costs associated with falls among the elderly are projected to rise from an estimated \$20 billion in 2000 to a staggering \$54.9 billion by 2020, as reported by the Centers for Disease Control and Prevention (CDC).

The consequences of falls extend far beyond immediate injuries, as many elderly fall victims cannot regain their footing independently, requiring assistance that may be delayed. Shockingly, individuals can wait an average of 10 minutes or longer, with 3\% enduring an hour or more of helplessness before receiving aid. Prolonged immobility during these critical periods often leads to further health complications, hospitalizations, institutionalization, and increased morbidity and mortality rates. Given these alarming statistics, it is crucial to implement comprehensive prevention strategies that combine education, training, secure environments, and innovative research initiatives supported by effective policy interventions to mitigate fall risks.

Existing literature offers numerous strategies to reduce fatal falls and improve the response times of medical and nursing staff. However, many of these solutions face high costs, complex implementations, or privacy limitations. To address these obstacles, we present a cost-effective embedded fall detection device that leverages accelerometers and gyroscopes, providing an unparalleled user-friendly experience. Our research endeavours encompass pioneering ideas, including developing a wearable node integrating fall detection, victim localization, and staff notification functions into a single device. Additionally, we introduce a robust and reliable Long Short-Term Memory (LSTM) model meticulously compared to conventional machine learning (ML) algorithms to enhance the accuracy and efficiency of fall detection. Complementing these advancements, we have designed an intuitive Android application to assist caregivers in providing care and support.

By merging cutting-edge technologies, cost-effective design, and a holistic perspective on fall detection and prevention, our research aims to alleviate the burden of falls, empower caregivers, and enhance the overall well-being of vulnerable individuals. Our comprehensive approach addresses the pressing need for effective fall prevention strategies, offering promising avenues to mitigate risks and improve the outcomes for fall victims.
\section{Literature Survey}
This chapter focuses on the design of a fall detection system for monitoring geriatric healthcare and detecting falls. Figure 1 demonstrates various Fall Detection systems.
\begin{figure}[!h]
  \centering
  \begin{tikzpicture}[node distance=2cm]
    \node[rectangle, draw] (fall) {Fall Detection System};
    
    \node[rectangle, draw, below left of=fall, node distance=2cm] (vision) {Vision Based};
    
    \node[rectangle, draw, below right of=fall, node distance=2cm] (ambient) {Ambient Based};
    
    \node[rectangle, draw, below of=fall, node distance=2cm] (wearable) {Wearable Based};
    
    \draw[-] (fall) -- (vision);
    \draw[-] (fall) -- (ambient);
    \draw[-] (fall) -- (wearable);
    
\end{tikzpicture}
  \caption{Various types of a Fall Detection System}
\end{figure}
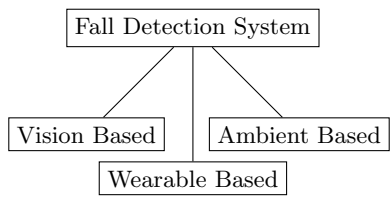
\begin{itemize}
    \item \textbf{Vision-based System}
\end{itemize}
Anishchenko \cite{anishchenko2019fall} deep learning and transfer learning methodologies on real-world surveillance camera data to identify instances of falls to address the limitations associated with artificially generated datasets obtained from controlled scenarios. Bhandari et al.\cite{sase2018human} utilize a three-step approach to detect falls in video frames, comprising identifying interest points through the Shi-Tomasi algorithm, determining the inter-point distances by calculating optical flow with the Lucas-Kanade algorithm and estimating motion speed and direction to determine the occurrence of falls. Ogden Kwolek.\cite{kwolek2013fall} analyzed Kinect camera feeds and utilized point cloud images to detect falls.

Vision-based systems in healthcare offer precise information through images or video feeds, aiding remote caregivers and enabling early detection of health issues. However, they can be costly to implement and maintain, time-consuming to process, and raise privacy concerns.

\begin{itemize}
    \item \textbf{Ambient-based System}
\end{itemize}
Taramasco et al.\cite{taramasco2018novel} describes a fall classification system that utilizes low-resolution thermal sensors placed at two horizontal planes near the floor. The results revealed that the Bi-LSTM model achieved a high accuracy of 93\%, outperforming the other RNN models.

This ambient system offers several advantages, including its comfortable nature as it does not require users to wear any devices or sensors. It enables continuous monitoring, even without the user wearing any sensors. In case of disadvantages, False alarms may also occur if certain activities are misinterpreted as falls, such as when the user is sitting or lying down.
\begin{itemize}
    \item \textbf{Wearable-based System}
\end{itemize}
Kaewkannate and Kim \cite{kaewkannate2016comparison} comprehensively analyze four wearable devices designed in a wristband style. Their evaluation thoroughly compares each device's various features and costs. On the other hand, the power consumption of wearable devices is highly dependent on several factors. He et al.\cite{ranakoti2019human} combined tri-axial accelerometers with gyroscopes and magnetometers to capture a comprehensive range of motion data. Their wearable device demonstrated enhanced performance in detecting falls and differentiating them from everyday activities.

Wearable fall detection devices offer increased safety and improved response time, particularly for individuals at risk of falls. They have a user-friendly design and are often more cost-effective than alternative caregiving options. However, these devices may be prone to false alarms and have limited effectiveness in detecting certain types of falls, posing user challenges.

\section{Preliminaries}
\subsection{Problem Statement}
The growing concern about falls among the elderly necessitates the development of accurate and practical fall detection systems. Vision-based and ambient-based approaches have limitations in terms of accuracy and practicality. Wearable fall detection systems offer continuous and unobtrusive monitoring, overcoming the boundaries of existing methods. However, achieving high recall while maintaining acceptable precision is a challenge. This paper aims to develop a wearable fall detection system that prioritizes high memory using sensors and LSTM models. By leveraging wearable sensor technology and LSTM models, this system aims to enhance the safety and well-being of vulnerable individuals.

\subsection{Relevance of LSTM in Fall Detection}
LSTM\cite{hochreiter1997long} is a valuable approach in fall detection, addressing precision and recall challenges. Recall is crucial in fall detection to minimize false negatives and ensure timely assistance. LSTM detects sudden sit and fall events by capturing their temporal dynamics and leveraging its memory component. Compared to methods like MLP, LSTM's ability to capture long-term dependencies enhances its effectiveness in recognizing fall patterns. Utilizing LSTM in fall detection systems can significantly improve the safety and well-being of individuals requiring monitoring.

\subsection{Sensor}
\begin{figure}[hbtp]
  \centering
  \includegraphics[width=0.2\textwidth]{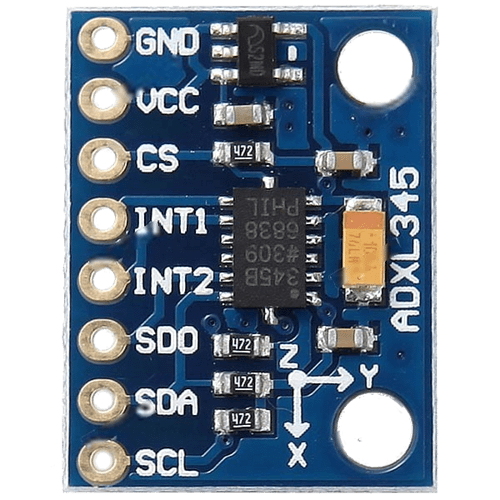}
  \caption{ADXL345 Sensor }
  \label{fig:example}
\end{figure}
The ADXL345 sensor is reliable and versatile for applications requiring precise motion sensing and acceleration measurement. It offers exceptional performance with low power consumption, high resolution, and programmable range options. With its integrated 3-axis gyroscope functionality, the ADXL345 provides comprehensive motion-sensing capabilities. Its ability to accurately measure changes in acceleration makes it well-suited for fall detection applications. The sensor's low power consumption and compact size make it ideal for integration into wearable devices. The programmable features of the ADXL345 enable customization for optimizing fall detection accuracy and minimizing false alarms. 

\section{Proposed Methodology}
\subsection{Overview of Hardware system}
Figure 3 provides an overview of the total system, encompassing power management, data acquisition, and communication in a mobile device. Key components include the Power Manager Unit for efficient power distribution, an accelerometer and gyroscope for accurate motion sensing, Raspberry Pi 3 B+ as the central processing unit, GPS and SIM908 modules for precise positioning, a GSM module for network communication, and a user device for interaction with the system.
\tikzstyle{block} = [draw, rectangle, minimum height=3em, minimum width=4em]
\tikzstyle{arrow} = [thick,->,>=stealth]
\begin{figure}
    \centering
    \begin{tikzpicture}[auto, node distance=1.5cm,>=latex]
        \node [block] (pmu) {Power Manager Unit};
        \node [block, below of=pmu] (imu) {ADXL345};
        \node [block, right of=pmu, node distance=3.5cm] (rpi) {Raspberry Pi 3 B+};
        \node [block, below of=rpi] (gps) {GPS Antenna , GPS Module};
        \node [block, below of=gps] (gsm) {GSM Module , GSM Antenna};
        \node [block, below of=gsm] (device) {User Device};
        \draw [->] (pmu) -- (imu);
        \draw [->] (pmu) -- (rpi);
        \draw [->] (imu) -- (rpi);
        \draw [->] (rpi) -- (gps);
        \draw [->] (gps) -- (gsm);
        \draw [->] (gsm) -- (gps);
        \draw [->] (gsm) -- (device);

    \end{tikzpicture}
    \caption{Overview of the Hardware system}
    \label{fig:System overview}
\end{figure}
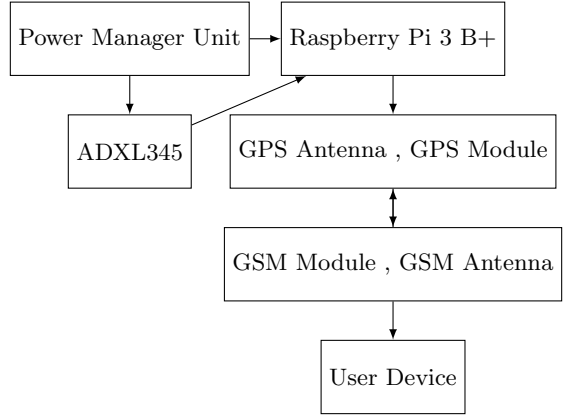

\subsection{Data-Preprossessing}
Data preprocessing enhances fall detection system performance by addressing noise, drift, and artifacts in accelerometer and gyroscope sensor data. Filtering techniques like low-pass and median filters reduce noise and outliers. Normalization scales data to a standardized range, ensuring consistent comparisons across sensors or individuals. Feature extraction identifies relevant patterns, such as statistical measures or frequency-domain features, facilitating accurate fall detection. The Butterworth filter provides a smoothing effect, highlighting significant variations in acceleration or angular velocity associated with falls: filter order and cutoff frequency selection balance precision, computational complexity, and phase distortion. Min-Max normalization eliminates biases and scaling effects, improving accuracy and robustness. Z-score normalization is an alternative technique. The choice of preprocessing methods depends on system requirements and sensor data characteristics.
\begin{figure}[hbtp]
  \centering
    {\includegraphics[width=0.35\textwidth]{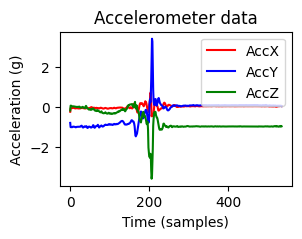}\label{fig:image1}}
  {\includegraphics[width=0.35\textwidth]{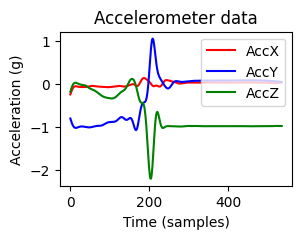}\label{fig:image2}}
  \caption{Graphical representation of raw and filtered data}
  \label{fig:four_images}
\end{figure}
Figure 4 visually represents accelerometer data for a forward fall, showcasing both the raw and filtered data.

\subsection{Proposed Model}
In our research, we utilized a Long Short-Term Memory (LSTM) model, a type of recurrent neural network, for fall detection. There are two LSTM layers with 64 and 32 memory units each. This design effectively captured temporal dependencies and patterns in the accelerometer and gyroscope data, enabling accurate classification of falls. The input data for the LSTM model included accelerometer and gyroscope measurements from three axes, providing a comprehensive representation of motion data. The model processed sequences of 50-time steps, each with six features (3-axis accelerometer and 3-axis gyroscope measurements).To prevent overfitting, Dropout layers were incorporated after the LSTM layers. The LSTM model employed an output layer with two units representing fall and non-fall classes. A softmax activation function generated probabilities for each category, allowing the model to estimate the likelihood of input sequences belonging to each class. The model was trained using the categorical cross-entropy loss function to minimize the discrepancy between predicted and actual class labels. The Adam optimizer, known for adapting learning rates for individual model parameters, updated the model's weights based on computed gradients. Default parameters were employed, and a batch size of 32 was used to facilitate efficient parameter updates and speed up model convergence.

\subsection{ Applied Weight Pruning Techniques}
\begin{figure}[hbtp]
  \centering
  \resizebox{0.3\textwidth}{!}{
  \begin{tikzpicture}[node distance=2cm, scale=0.8]
    \node[draw, rectangle] (input) {Input};
    \node[draw, rectangle, right of=input] (lstm) {LSTM};
    \node[draw, rectangle, right of=lstm] (output) {Output};
    \node[draw, rectangle, below of=lstm] (pruning) {Pruning};
    \node[draw, rectangle, below of=pruning] (compressed) {Compressed Model};
    
    \draw[->] (input) -- (lstm);
    \draw[->] (lstm) -- (output);
    \draw[->] (lstm) -- (pruning);
    \draw[->] (pruning) -- (compressed);
    \draw[->] (compressed) -- (output);
  \end{tikzpicture}
  }
  \caption{Pipeline of Weight Pruning in LSTM}
  \label{fig:pruning_pipeline}
\end{figure}
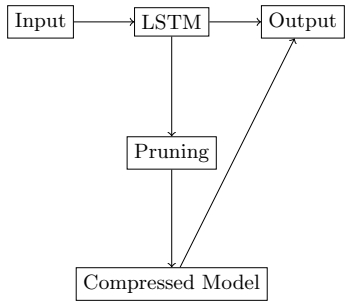
To optimize the LSTM model, weight pruning techniques were applied. Weight pruning involves selectively removing less significant connections or weights from the network while preserving accuracy. Figure 5 represents the pipeline of weight pruning in LSTM. The goal was to achieve a 10 to 30\% pruning sparsity rate, meaning the percentage of pruned weights.
Weight pruning offers benefits beyond model compression. It improves computational efficiency by reducing computations during inference. The reduced model size enables easier deployment on resource-constrained devices. Weight pruning contributes to developing efficient and lightweight fall detection systems suitable for real-world applications.
\section{Result}
\subsection{Dataset}
\begin{table}[h]
\scriptsize
\centering
\begin{tabular}{ |c||c|c| }
\hline
\multicolumn{3}{|c|}{ADL Activities} \\
\hline
Description & Notations & Data points\\
\hline
Walking & NF1 & 980\\
Sitting & NF2 & 1010\\
Lying & NF3 & 970\\
Running & NF4 & 1200\\
Sudden Sit & NF4 & 1100\\
Sudden Standing & NF5 & 1220\\
\hline
\end{tabular}
\begin{tabular}{ |c||c|c| }
\hline
\multicolumn{3}{|c|}{FALL Activities} \\
\hline
Description & Notations & Data points\\
\hline
Fall Forward & F1 & 1200\\
Fall Backward & F2 & 990\\
Fall Left & F3 & 1000\\
Fall Right & F4 & 1100\\
\hline

\end{tabular}
\caption{\label{demo-table} Description of six types of ADL and four types of fall}
\end{table}
Table 1 represents the collected real-time data to differentiate between falls and activities of daily living (ADLs), a group of six individuals (three women and three men, aged 30 to 60) participated. Their heights ranged from 160 cm to 185 cm, and their weights varied from 50 kg to 85 kg. To achieve this goal, the participants performed six different types of ADLs and four types of falls under controlled conditions using a mattress with a thickness of 20 cm. A total of 10,770 data points were collected during the study, with 6,480 data points corresponding to ADLs and 4,290 data points corresponding to falls. The data points were collected from the participants while performing the designated activities, with variations in activity frequencies among participants. Participants A, B, C, D, and E (all aged 30) completed ten exercises each. In contrast, participants E, F (aged 55), G, H (aged 58), and I (aged 60) performed a different number of activities based on individual capabilities or study requirements.

\subsection{Experimental Setup}
 The research utilized Anaconda with Keras and TensorFlow on Windows 10 for training LSTM models. Testing was conducted on a Raspberry Pi 3 B+ in a Linux environment. Jupyter Notebook was used for code execution and result analysis. This approach allowed for evaluating model performance on different platforms and assessing real-world feasibility.

\subsection{Confusion Matrix}
\begin{figure}[hbtp]
  \centering
  \includegraphics[width=0.4\textwidth]{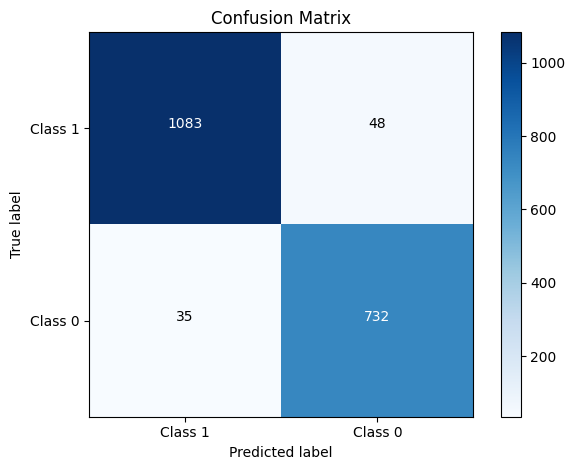}
  \caption{Confusion Matrix}
  \label{fig:image_label}
\end{figure}
Figure 6 represents the confusion matrix for the fall detection system can be summarized as follows:

The system achieved the following in the case of non-fall instances (Class 0).1083 true negatives (TN), correctly classifying them as non-fall instances.48 false positives (FP), incorrectly classifying them as fall instances. For fall instances (Class 1), the system achieved the following. Seven hundred thirty-two true positives (TP), correctly identifying them as fall instances. Thirty-five false negatives (FN), incorrectly classifying them as non-fall instances.

This information provides valuable insights into the system's classification performance for fall and non-fall instances.

\subsection{Classification Report}
Class 0 (non-fall) instances were classified with 97\% precision and 96\% recall, resulting in an F1-score of 96\%. For class 1 (fall) instances, the precision was 94\%, the recall was 95\%, and the F1-score was 95\%. These metrics demonstrate the system's high accuracy in identifying non-fall and fall instances. Figure 7 and 8 represents train test accuracy and train test loss.

\begin{figure}[h]
  \begin{minipage}[b]{0.3\linewidth}
    \centering
    \includegraphics[width=\linewidth]{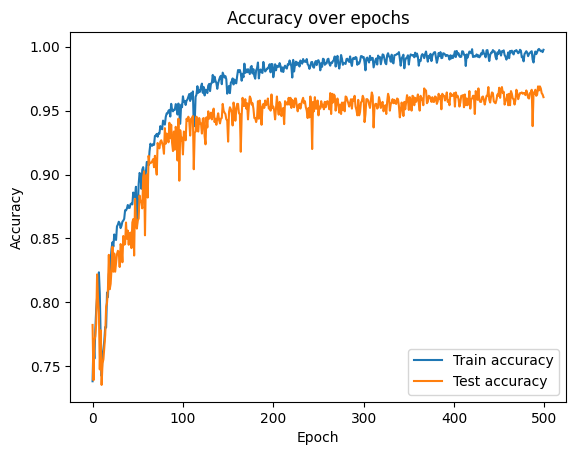}
    \caption{Train and Test accuracy}
    \label{fig:image1}
  \end{minipage}
  \hfill
  \begin{minipage}[b]{0.3\linewidth}
    \centering
    \includegraphics[width=\linewidth]{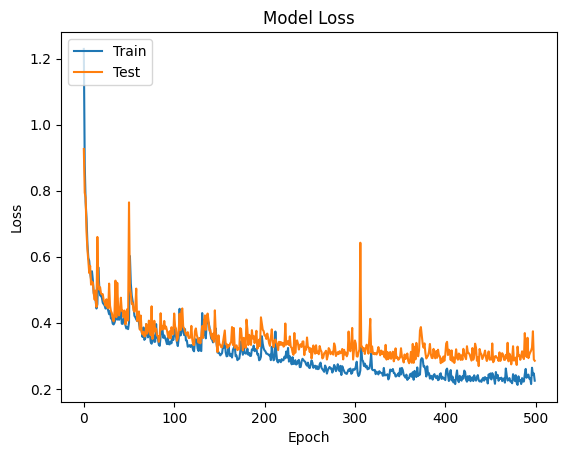}
    \caption{Train and Test Loss}
    \label{fig:image2}
  \end{minipage}
  \hfill
  \begin{minipage}[b]{0.3\linewidth}
    \centering
    \includegraphics[width=\linewidth]{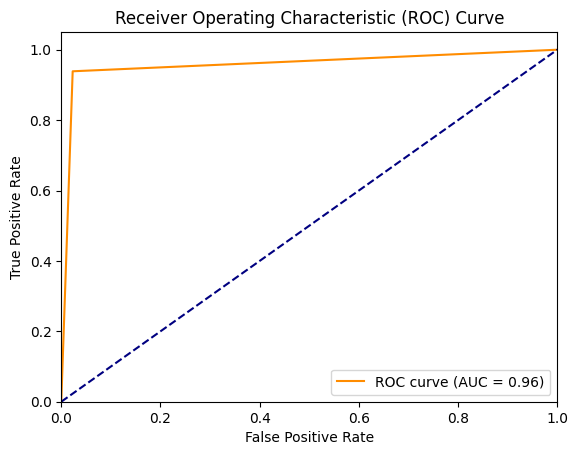}
    \caption{ROC curve}
    \label{fig:image3}
  \end{minipage}
  \label{fig:three_images}
\end{figure}

\subsection{Receiver Operating Characteristic (ROC) Curve}
Figure 9 displays the Receiver Operating Characteristic (ROC)\cite{fawcett2006introduction} curve, visually representing the fall detection system's performance. The ROC curve showcases the trade-off between sensitivity (true positive rate) and specificity (1 - false positive rate) at various classification thresholds.
The Area quantifies the performance of the fall detection system Under the ROC Curve (AUC). The AUC value, calculated as 0.96 in this case, accurately measures the system's ability to differentiate between fall and non-fall instances. A higher AUC value signifies a more substantial discriminatory power of the system.

\section{Comparison of different Models and Proposed LSTM}
The comparison of different models involved assessing their performance based on various metrics such as accuracy, precision, and recall. Table 2 presents valuable insights into the models' performance in accurately classifying fall and non-fall instances while effectively reducing false positives and negatives.
\begin{table}[h]
\centering
\resizebox{0.8\linewidth}{!}{%
\begin{tabular}{|l|c|c|c|c|}
\hline
\textbf{References} & \textbf{Sensors} & \textbf{Algorithm} & \textbf{Sensitivity} & \textbf{Specificity} \\
\hline
Shi et al. \cite{shi2012fall} & A & SVM & 90.0 & 95.7 \\
Aguiar et al. \cite{aguiar2014accelerometer} & A & Dec. Tree & 94.0 & 90.2 \\
Helmy and Helmy \cite{helmy2015seizario} & A & Thresholds & 95.0 & 90.0 \\
Tran et al. \cite{tran2017new} & A & Perceptron & 60.4 & 94.8 \\
Proposed Model & A, G & LSTM & 95 & 96 \\
\hline
\end{tabular}
}
\caption{Fall Detection Systems Comparison}
\label{tab:comparison}
\end{table}

(A - Accelerometer, G - Gyroscope, N/A - Not Applicable)

\section{Conclusion and Future Work}
This research addressed the critical issue of fall-related incidents by developing and evaluating a fall detection system using LSTM. The system achieved high accuracy, precision, and recall, effectively minimizing false negatives. The LSTM-based system demonstrated competitive accuracy, sensitivity, and specificity performance compared to previous models. Future research directions include model optimization, sensor integration, real-time implementation, dataset expansion, and validation in real-world scenarios. Overall, this thesis improves the safety and quality of life for individuals at risk of falls.
Future research should focus on utilizing more lightweight sensors and incorporating additional features to improve the accuracy of the fall detection system.

\bibliographystyle{splncs04}
\bibliography{cit}

\begin{thebibliography}{10}
\providecommand{\url}[1]{\texttt{#1}}
\providecommand{\urlprefix}{URL }
\providecommand{\doi}[1]{https://doi.org/#1}

\bibitem{aguiar2014accelerometer}
Aguiar, B., Rocha, T., Silva, J., Sousa, I.: Accelerometer-based fall detection for smartphones. In: 2014 IEEE International Symposium on Medical Measurements and Applications (MeMeA). pp.~1--6. IEEE (2014)

\bibitem{anishchenko2019fall}
Anishchenko, L., Zhuravlev, A., Chizh, M.: Fall detection using multiple bioradars and convolutional neural networks. Sensors  \textbf{19}(24), ~5569 (2019)

\bibitem{fawcett2006introduction}
Fawcett, T.: An introduction to roc analysis. Pattern recognition letters  \textbf{27}(8),  861--874 (2006)

\bibitem{helmy2015seizario}
Helmy, A., Helmy, A.: Seizario: Novel mobile algorithms for seizure and fall detection. In: 2015 IEEE Globecom Workshops (GC Wkshps). pp.~1--6. IEEE (2015)

\bibitem{hochreiter1997long}
Hochreiter, S., Schmidhuber, J.: Long short-term memory. Neural computation  \textbf{9}(8),  1735--1780 (1997)

\bibitem{kaewkannate2016comparison}
Kaewkannate, K., Kim, S.: A comparison of wearable fitness devices. BMC public health  \textbf{16},  1--16 (2016)

\bibitem{kwolek2013fall}
Kwolek, B., Kepski, M.: Fall detection using kinect sensor and fall energy image. In: Hybrid Artificial Intelligent Systems: 8th International Conference, HAIS 2013, Salamanca, Spain, September 11-13, 2013. Proceedings 8. pp. 294--303. Springer (2013)

\bibitem{ranakoti2019human}
Ranakoti, S., Arora, S., Chaudhary, S., Beetan, S., Sandhu, A.S., Khandnor, P., Saini, P.: Human fall detection system over imu sensors using triaxial accelerometer. In: Computational Intelligence: Theories, Applications and Future Directions-Volume I: ICCI-2017. pp. 495--507. Springer (2019)

\bibitem{sase2018human}
Sase, P.S., Bhandari, S.H.: Human fall detection using depth videos. In: 2018 5th International Conference on Signal Processing and Integrated Networks (SPIN). pp. 546--549. IEEE (2018)

\bibitem{shi2012fall}
Shi, Y., Shi, Y., Wang, X.: Fall detection on mobile phones using features from a five-phase model. In: 2012 9th International Conference on Ubiquitous Intelligence and Computing and 9th International Conference on Autonomic and Trusted Computing. pp. 951--956. IEEE (2012)

\bibitem{taramasco2018novel}
Taramasco, C., Rodenas, T., Martinez, F., Fuentes, P., Munoz, R., Olivares, R., De~Albuquerque, V.H.C., Demongeot, J.: A novel monitoring system for fall detection in older people. Ieee Access  \textbf{6},  43563--43574 (2018)

\bibitem{tran2017new}
Tran, H.A., Ngo, Q.T., Tong, V.: A new fall detection system on android smartphone: Application to a sdn-based iot system. In: 2017 9th International Conference on Knowledge and Systems Engineering (KSE). pp.~1--6. IEEE (2017)

\end{thebibliography}

\end{document}